\documentclass[aps,preprint,nofootinbib,preprintnumbers,eqsecnum,superscriptaddress]{revtex4}
\pdfoutput=1

\usepackage{wrapfig}

\usepackage{color}

\usepackage[
      colorlinks=true,
      linkcolor=blue,
      urlcolor=blue,
      filecolor=black,
      citecolor=red,
      pdfstartview=FitV,
      pdftitle={},
        pdfauthor={Alejandra Castro, Stephane Detournay, Nabil Iqbal, Eric Perlmutter},
        pdfsubject={},
        pdfkeywords={},
        pdfpagemode=None,
        bookmarksopen=true
      ]{hyperref}

\usepackage[font=footnotesize,labelfont=bf,justification=centerlast,width=.94\textwidth]{caption}

\usepackage[normalem]{ulem}
\usepackage{amsmath}
\usepackage{enumerate}
\usepackage{amsfonts}
\usepackage{yfonts}

\usepackage{subfigure}
\usepackage{psfrag}

\usepackage{epsfig}
\usepackage[latin1]{inputenc}
\usepackage{float}
\usepackage{graphicx}
\usepackage{cancel}
\usepackage{mathrsfs}
\usepackage{amssymb}
\usepackage{amsfonts}
\usepackage{amsmath}
\usepackage{slashed}

\usepackage{graphicx}
\usepackage{bm}

\newcommand{\be}{\begin{equation}}
\newcommand{\ee}{\end{equation}}
\newcommand{\bea}{\begin{eqnarray}}
\newcommand{\eea}{\end{eqnarray}}

\begin{document}

\title {Boson star with particle size effects}


\author{Claude Semay}
\affiliation{Service de Physique Nucl\'eaire et Subnucl\'eaire, Universit\'e de Mons, UMONS Research Institute for Complex Systems, Place du Parc 20, 7000 Mons, Belgium}

\begin{abstract}
\vskip 0.3in
A simple model to study boson stars is to consider these stellar objects as quantum systems of $N$ identical self-gravitating particles within a non-relativistic framework. Some results obtained with point-like particles are recalled as well as the validity limits of this model. Approximate analytical calculations are performed using envelope theory for a truncated Coulomb-like potential simulating a particle size. If the boson mass is sufficiently small, the description of small mass boson stars is possible within non-relativistic formalism. The mass and radius of these stellar objects are strongly dependent on the value of the truncation parameter. 
\end{abstract}

\vfill


\maketitle

\tableofcontents

\section{Introduction}
\label{sect:intro}

A boson star is a stellar object made of bosons, contrary to conventional stars which are formed of fermions \cite{jetz92}. At present, there is no experimental evidence that such a star exists. However, a pair of co-orbiting boson stars could be detected by their emission of gravitational radiation \cite{pale08}. The framework of general relativity is certainly the best one to study such a hypothetical object \cite{jetz92,brih15}, but interesting results can be obtained in a simpler model, in which the boson star is considered as a quantum system of $N$ identical self-gravitating particles within a non-relativistic framework \cite{jetz92}. The corresponding Hamiltonian is then 
\begin{equation}
\label{HBS1}
H=\sum_{i=1}^N \frac{\bm p_i^2}{2 m} - \sum_{i< j=2}^N \frac{G m^2}{|\bm r_i - \bm r_j|},
\end{equation}
where $m$ is the mass of the boson and $G$ is the gravitational constant. Approximate solutions for Hamiltonian~(\ref{HBS1}) have already been computed \cite{jetz92,basd90}. In this paper, these results are recalled and commented on, and the effect of a possible size for the boson is studied. In order to simulate such a phenomenon, the usual interaction can be replaced by a truncated Coulomb-like potential 
\begin{equation}
\label{truncclp}
H_T=\sum_{i=1}^N \frac{\bm p_i^2}{2 m} - \sum_{i< j=2}^N \frac{G m^2}{|\bm r_i - \bm r_j|+b},
\end{equation}
where $b$ is linked to the size of the bosons. In a hard sphere picture, $b$ is the diameter of the particle and $r$ is the distance between their surfaces. This potential, which avoids singularity at $r=0$, is widely used in several domains of physics \cite{sing85,fern91}. The boson size could be due to the compound nature of the boson or to the existence of a natural minimal length. More modestly, this size can be considered as a means to simulate unknown effects aimed at reducing the singularity of the gravitational interaction. 

The technique to solve $N$-body systems is presented in Section~\ref{sect:et}. The results obtained for the Hamiltonian~(\ref{HBS1}) are recalled in Section~\ref{sect:point}. The results obtained for the truncated Coulomb-like potential~(\ref{truncclp}) are computed in Section~\ref{sect:size}. Finally, some concluding remarks are given in Section~\ref{sect:rem} with possible extensions. In this paper, a space with $D$ dimensions is considered, assuming that the gravitational interaction is always given by a (truncated) Coulomb-like potential. 

\section{Envelope theory}
\label{sect:et}

Among the various techniques aimed at solving the quantum problem of $N$-body systems, envelope theory (ET), which was developed several years ago, is particularly interesting. It is quite easy to implement and can yield upper or lower bounds \cite{hall80,hall04}. The idea is to find a known envelope potential and/or a known envelope kinetic part for the Hamiltonian under consideration. This procedure has been rediscovered and extended in another way under the name of `auxiliary field method' \cite{silv10}. A practical form for the equations of ET is given in \cite{sema13a} for quite general Hamiltonians. It has been checked that ET can give reasonable accuracy for the upper or lower bounds of various systems \cite{sema15a}. 

Let us consider a general Hamiltonian, in a $D$ dimensional space ($D \ge 2$), for $N$ identical particles 
\begin{equation}
\label{HNb}
H=\sum_{i=1}^N T(|\bm p_i|) + \sum_{i=1}^N U\left(|\bm r_i - \bm R|\right) + \sum_{i< j=2}^N V\left(|\bm r_i - \bm r_j|\right).
\end{equation}
$T$ is a kinetic energy, $U$ a one-body interaction, $V$ a two-body  potential and $\bm R = \frac{1}{N}\sum_{i=1}^N \bm r_i$ is the centre of mass position. In the framework of ET, an approximate eigenvalue $E$ is given by the following set of equations for a completely (anti)symmetrised state and the centre of mass motion removed ($\sum_{i=1}^N \bm p_i = \bm 0$) \cite{sema13a}
\begin{eqnarray}
\label{AFM1N}
&&E=N\, T(p_0) + N\, U \left( \frac{r_0}{N} \right) + C_N\, V \left( \frac{r_0}{\sqrt{C_N}} \right), \\
\label{AFM2N}
&&r_0\, p_0=Q, \\
\label{AFM3N}
&&N\, p_0\, T'(p_0) =  r_0\, U' \left( \frac{r_0}{N} \right) + \sqrt{C_N}\, r_0\, V' \left( \frac{r_0}{\sqrt{C_N}} \right),
\end{eqnarray}
where $W'(x)=dW/dx$, $C_N=N(N-1)/2$ is the number of particle pairs. 
\begin{equation}
\label{QN}
Q = \sum_{i=1}^{N-1} (\phi\, n_i + l_i) + (N - 1)\frac{D+\phi-2}{2}
\end{equation}
is a global quantum number. Following the forms of $T$, $U$ and $V$, the approximate value $E$ can have a variational character \cite{sema13a}. The method is easy to implement since the solution can be obtained simply through a transcendental equation. In the original method, $\phi = 2$, which corresponds to the global quantum number of $N-1$ identical harmonic oscillators. It has been shown that allowing variations of $\phi$ can improve the accuracy of the approximate eigenvalues \cite{sema15a}. An estimation of a relevant value for $\phi$ can actually be computed by comparing ET with the dominantly orbital state method \cite{sema15b}. This idea was inspired by \cite{loba09}, in which a universal effective quantum number for centrally symmetric 2-body systems is proposed. 

Using the following relations \cite{sema15b},
\begin{eqnarray}
\label{p0}
p_0^2 &=& \frac{1}{N}\left\langle \sum_{i=1}^N \bm p_i^2 \right\rangle, \\
\label{r0}
r_0^2 &=& N \left\langle \sum_{i=1}^N (\bm r_i - \bm R)^2 \right\rangle =
\left\langle \sum_{i<j=2}^N (\bm r_i - \bm r_j)^2 \right\rangle, 
\end{eqnarray}
it is easy to compute the mean values of the momentum $p_*$ of a particle, of the radius $R_*$ of the system and of its mass $M_*$:
\begin{equation}
\label{star}
p_* = p_0, \quad R_* = \frac{r_0}{N}, \quad M_* = Nm+E.
\end{equation}
In the following, only the ground state will be considered, that is to say $Q=(N - 1)(D+\phi-2)/2$. 

\section{Point-like particles}
\label{sect:point}

The ground state solution of the ET equations for (\ref{HBS1}) is given by
\begin{eqnarray}
\label{r0pp}
r_0&=& \sqrt{\frac{N-1}{2 N}} \frac{(D+\phi-2)^2}{G m^3},\\
\label{Epp}
E&=& - \frac{N^2(N-1)}{4} \frac{G^2 m^5}{(D+\phi-2)^2}.
\end{eqnarray}
With $\phi=2$, (\ref{Epp}) is an upper bound. With $\phi=1$, the value given by the procedure of \cite{sema15b}, (\ref{Epp}) is the exact result for $N=2$ and a lower bound for $N>2$ \cite{sema15a,sema15b}. The quality of this last bound is similar to the ones obtained with other methods \cite{jetz92,basd90}. For large values of $N$, looking at the energy ($E \propto N^3$) and the radius ($R_* \propto N^{-1}$) behaviours, it can be expected that the non-relativistic model will break down if the number of particles increases too much. In fact, three limits can be defined:
\begin{description}
  \item[Momentum limit (ML)] The mean speed of a boson inside the star must be much lower than the speed of light. This is equivalent to $p_* \ll m$. So, it is possible to define a number of particles $N_{\textrm{ML}}$, such that the last inequality can be written $N \ll N_{\textrm{ML}}$. 
  \item[Energy limit (EL)] The binding energy of the star $E$ must be much lower than the rest energy of the bosons, that is to say $|E| \ll Nm$. As for the previous point, a condition $N \ll N_{\textrm{EL}}$ can be defined. 
  \item[Radius limit (RL)] The radius of the star $R_*$ must be much greater than the Schwarzschild radius of the star $R_S=2GM_*$. If the energy limit is satisfied, this is equivalent to $R_* \gg 2GNm > 2GM_*$. Again, an equivalent condition $N \ll N_{\textrm{RL}}$ can be defined.
\end{description}
It can be assumed that $m\ll m_{\textrm{Pl}}$, where $m_{\textrm{Pl}}=1/\sqrt{G}=1.2\times 10^{19}$~GeV is the Planck mass. Then, introducing the critical number,
\begin{equation}
\label{nc}
N_c = \frac{D+\phi-2}{G m^2}=(D+\phi-2)\frac{m_{\textrm{Pl}}^2}{m^2},
\end{equation}
it is easy to show that 
\begin{equation}
\label{nlimpp}
N_{\textrm{ML}}\approx \sqrt{2}N_c, \quad N_{\textrm{EL}}\approx 2 N_c, \quad N_{\textrm{RL}}\approx 2^{-3/4}N_c.
\end{equation}
These three limiting numbers are essentially the same. 

The boson star mass increases quasi-linearly with $N$, reaches a maximum, and then undergoes a brutal collapse. The maximal mass for the boson star is 
\begin{equation}
\label{maxmbs}
M_*^{\textrm{max}}\approx \frac{4(D+\phi-2)}{3^{3/2}} \frac{m_{\textrm{Pl}}^2}{m}.
\end{equation}
This mass corresponds to $N \approx 2N_c/\sqrt{3}$ and $E\approx -M_*^{\textrm{max}}/3$. So the non-relativistic theory breaks down before this maximum is reached. The number of protons in the Sun is approximatively $10^{57}$. $N_c$ reaches this value if $m\approx 0.5$~eV ($D=3$, $\phi=1$). In this case, $M_*^{\textrm{max}}\approx 8\times 10^{20}$~kg, which is around $10\ 000$ times smaller than the Earth's mass. 

\section{Finite size effects}
\label{sect:size}

The computation of the ground state solution of the ET equations for (\ref{truncclp}) is more involved, but the solution is still in a closed-form. Let us define
\begin{eqnarray}
\label{ga}
A&=&\frac{(D+\phi-2)^2}{N G m^3 b}, \\
\label{x0}
x_0 &=& \frac{\sqrt{A(A+6)}}{3}F_{-}\left( \frac{2 A^2+18 A+27}{2 A^{1/2}(A+6)^{3/2}} \right)+\frac{A}{3}
\end{eqnarray}
where $F_\pm(Y)$ is the only positive root of the cubic equation $x^3 \pm 3x - 2Y =0$ with $Y \ge 0$ \cite{silv10}.\footnote{$F_\pm(Y) = \left(Y + \sqrt{Y^2\pm 1} \right)^{1/3} \mp \left(Y + \sqrt{ Y^2\pm1} \right)^{-1/3}$ for $Y\ge 1$ and $F_{-}(Y) = 2 \cos\left( \frac{1}{3} \arccos Y \right)$ for $Y < 1$.} This gives 
\begin{eqnarray}
\label{r0tc}
r_0&=& \sqrt{\frac{N(N-1)}{2}}\, b\, x_0,\\
\label{Etcp}
E&=& - \frac{N(N-1)}{2}\frac{G m^2}{b} \frac{x_0 + 2}{2(x_0 +1)^2}.
\end{eqnarray}
The procedure of \cite{sema15b} gives 
\begin{equation}
\label{phitrunc}
\phi=\sqrt{\frac{\tilde x_0+3}{\tilde x_0+1}},
\end{equation}
where $\tilde x_0$ is given by (\ref{x0}) in which $A|_{\phi=0}=(D-2)^2/(N G m^3 b)$. For $\phi=2$, (\ref{Etcp}) is an upper bound. For other values of $\phi$, (\ref{Etcp}) has no defined variational character. 

The radius and mass of the star are complicated functions of $m$ and $b$. Numerical tests show that boson star mass increases quasi-linearly with $N$, reaches a maximum, and then undergoes a brutal collapse, as in the $b=0$ case. The behaviour of the solutions is ruled by the parameter $A$. For large $N$, such that 
\begin{equation}
\label{ainv}
N \gg (D+\phi-2)^2 \frac{m_{\textrm{Pl}}^3}{m^3} \frac{l_{\textrm{Pl}}}{b},
\end{equation}
where $l_{\textrm{Pl}}=1/m_{\textrm{Pl}}$ is the Planck length, 
\begin{eqnarray}
\label{r0tcln}
r_0&=& \frac{N^{2/3}}{\sqrt{2}} \frac{b^{2/3}}{G^{1/3}m} (D+\phi-2)^{2/3} + O(N^{1/3}),\\
\label{Etcpln}
E&=& - \frac{N^2}{2}\frac{G m^2}{b} + O(N^{5/3}).
\end{eqnarray}
The convergence with $N$ is quite slow. The last equations can be used provided $b$ is larger than the Compton wavelength $2\pi/m$ of the boson. It can be noted that $E\propto N^2$ and $R_* \propto N^{-1/3}$, which is quite different from the $b=0$ case. The limiting numbers are then given by
\begin{equation}
\label{nlimpp}
N_{\textrm{ML}}\approx \frac{2^{3/2}b^2}{G(D+\phi-2)}, \quad N_{\textrm{EL}}\approx \frac{2 b}{G m}, \quad N_{\textrm{RL}}\approx \frac{b^{1/2}(D+\phi-2)^{1/2}}{2^{9/8}Gm^{3/2}}.
\end{equation}
The situation is quite different from the non-truncated potential, but a link exists between these numbers
\begin{equation}
\label{nlimlink}
N_{\textrm{EL}} = 2^{5/4} N_{\textrm{ML}}^{1/3} N_{\textrm{RL}}^{2/3}.
\end{equation}

The maximal mass for the boson star is then
\begin{equation}
\label{maxmbsb}
M_*^{\textrm{max}}\approx \frac{1}{2} b\,m_{\textrm{Pl}}^2.
\end{equation}
This mass, which is independent of $m$, is reached for $N \approx N_{\textrm{EL}}/2$. Again, the non-relativistic theory is questionable close to this maximum. Taking $m = 0.5$~eV and $b=10/m$ ($D=3$, $\phi$ given by (\ref{phitrunc})), $M_*^{\textrm{max}}\approx 2.6\times 10^{21}$~kg with formula~(\ref{maxmbsb}) and $4.1\times 10^{21}$~kg with formula~(\ref{Etcp}). It can be seen that the approximation~(\ref{maxmbsb}) is reasonable in this case, and that the results are quite different from the non-truncated potential.


\section{Concluding remarks}
\label{sect:rem}

With the genuine non-relativistic gravitational potential or with the truncated version simulating a boson size, the general non-relativistic behaviour of the boson star mass is the same: It increases quasi-linearly with $N$, reaches a maximum, and then undergoes a brutal collapse. Nevertheless, the characteristics of these stellar objects, mass and radius, can be quite different following the value of the truncation parameter $b$. Though  non-relativistic formalism is questionable for such systems, there is still some room for small mass boson stars (planetoids), if the boson mass is sufficiently small. 

In order to partly take into account relativistic effects, the non-relativistic kinematics $\bm p^2/(2 m)$ can be replaced by a relativistic one $\sqrt{\bm p^2+m^2}-m$ \cite{jetz92,rayn94}. This last operator is well-defined, but the corresponding model is not covariant. For point-like particles, the main effect is to lower the maximal boson star mass, but the collapse is still unavoidable for a large number of particles. This case can be analytically treated by ET, but no closed-form approximation can be obtained for the truncated Coulomb-like potential with semi-relativistic kinematics. One has to resort to numerical studies of the system~(\ref{AFM1N}-\ref{AFM3N}). 

\section*{Acknowledgments}

I would like to thank Professor Philippe Spindel, first as a teacher for his valuable lectures during my years as a student, second as a colleague for all the fruitful discussions we have had about various aspects of physics and life. I would also thank the organisers who gave me the opportunity to participate in this homage.

\providecommand{\href}[2]{#2}\begingroup\raggedright\endgroup


\begin{thebibliography}{10}

\bibliographystyle{utphys}


\bibitem{jetz92} P.~Jetzer, 
``{Boson Stars},''
\href{http://dx.doi.org/10.1016/0370-1573(92)90123-H}{{\em Phys. Rep.}
  {\bfseries 220} (1992) 163}.

\bibitem{pale08} C.~Palenzuela, L.~Lehner, S.L.~Liebling, 
``{Orbital dynamics of binary boson star systems},''
\href{http://dx.doi.org/10.1103/PhysRevD.77.044036}{{\em Phys. Rev. D}
  {\bfseries 77} (2008) 044036}.

\bibitem{brih15} Y.~Brihaye, B.~Hartmann, and J.~Riedel, 
``{Self-interacting boson stars with a single Killing vector field in anti-de Sitter space-time},''
\href{http://dx.doi.org/10.1103/PhysRevD.92.044049}{{\em Phys. Rev. D}
  {\bfseries 92} (2015) 044049}.

\bibitem{basd90} J.L.~Basdevant, A.~Martin, J.M.~Richard, 
``{Improved bounds on many-body hamiltonians (I). Self-gravitating bosons},''
\href{http://dx.doi.org/10.1016/0550-3213(90)90594-4}{{\em Nucl. Phys. B}
  {\bfseries 343} (1990) 60}.

\bibitem{sing85} D.~Singh, Y.P.~Varshni, R.~Dutt, 
``{Bound eigenstates for two truncated Coulomb potentials},''
\href{http://dx.doi.org/10.1103/PhysRevA.32.619}{{\em Phys. Rev. A}
  {\bfseries 32} (1985) 619}.

\bibitem{fern91} F.M.~Fern\'andez, 
``{Analytical bound eigenstates and eigenvalues of a truncated Coulomb potential},''
\href{http://dx.doi.org/10.1088/0305-4470/24/6/025}{{\em J. Phys. A}
  {\bfseries 24} (1991) 1351}.

\bibitem{hall80} R.L.~Hall, 
``{Energy trajectories for the $N$-boson problem by the method of potential envelopes},'' 
\href{http://dx.doi.org/10.1103/PhysRevD.22.2062}{{\em Phys. Rev. D} {\bfseries 22} (1980) 2062}.

\bibitem{hall04} R.L.~Hall, W.~Lucha, F.F.~Sch\"oberl,  
``{Relativistic $N$-boson systems bound by pair potentials $V(r_{ij}) = g(r^2_{ij})$},''
\href{http://dx.doi.org/10.1063/1.1767298}{{\em J. Math. Phys.}
  {\bfseries 45} (2004) 3086}.

\bibitem{silv10} B.~Silvestre-Brac, C.~Semay, F.~Buisseret, F.~Brau,
``{The quantum ${\cal N}$-body problem and the auxiliary field method},''
\href{http://dx.doi.org/10.1063/1.3340799}{{\em J. Math. Phys.}
  {\bfseries 51} (2010) 032104}.

\bibitem{sema13a} C.~Semay, C.~Roland, 
``{Approximate solutions for $N$-body Hamiltonians with identical particles in $D$ dimensions},''
\href{http://dx.doi.org/10.1016/j.rinp.2013.10.001}{{\em Results Phys.}
  {\bfseries 3} (2013) 231}.

\bibitem{sema15a} C. Semay, 
``{Numerical Tests of the Envelope Theory for Few-Boson Systems},''
\href{http://dx.doi.org/10.1007/s00601-015-0960-5}{{\em Few-Body Syst.}
  {\bfseries 56} (2015) 149}.

\bibitem{sema15b} C. Semay, 
``{Improvement of the envelope theory with the dominantly orbital state method},''
\href{http://dx.doi.org/10.1140/epjp/i2015-15156-7}{{\em Eur. Phys. J. Plus}
  {\bfseries 130} (2015) 156}.

\bibitem{loba09} A.A.~Lobashev, N.N.~Trunov, 
``{A universal effective quantum number for centrally symmetric problems},''
\href{http://dx.doi.org/10.1088/1751-8113/42/34/345202}{{\em J. Phys. A}
  {\bfseries 42} (2009) 345202}.

\bibitem{rayn94} J.C.~Raynal, S.M.~Roy, V.~Singh, A.~Martin, J.~Stubbe, 
``{The ``Herbst Hamiltonian'' and the mass of boson stars},''
\href{http://dx.doi.org/10.1016/0370-2693(94)90831-1}{{\em Phys. Lett. B}
  {\bfseries 320} (1994) 105}.

\end{thebibliography}

\end{document}